\documentclass{iopconfser}
\usepackage{amsmath,mathtools}
\usepackage{amssymb}
\usepackage{color,xcolor}
\usepackage{siunitx}
\usepackage[greek,english]{babel}
\usepackage{slashed}
\usepackage{appendix}
\begin{document}

\begin{flushleft}
  KCL-PH-TH/2024-{\bf 71}  
\end{flushleft}
\vspace{-2cm}

\title{Do we owe our existence to a Gravitational Anomaly? 
}

\author{Nick E. Mavromatos$^{1,2}$}

\affil{$^1$Physics Division, School of Applied Mathematical and Physical Sciences, National Technical University of Athens, Zografou Campus, Athens 15780, Greece} 
\vspace{0.1cm} 

\affil{$^2$ Theoretical Particle Physics and Cosmology Group, Physics Department, King’s College London, Strand, London WC2R 2LS, UK}

\email{nikolaos.mavromatos@kcl.ac.uk}

\begin{abstract}

I review a mechanism for a potential explanation of the dominance of matter  over antimatter in the Universe (``our existence''), based on a (3+1)-dimensional string-inspired cosmological model, characterised by (uncancelled) gravitational Chern-Simons (gCS) anomalies in the premordial epochs. The model is consistent with general covariance but entails spontaneous Lorentz (LV) and CPT Violation (CPTV), due to appropriate backgrounds of axion fields, which couple to the anomaly terms. Upon condensation of the gCS terms, due to primordial gravitational waves, these axions acquire a linear potential, which drives inflation of a Running-Vacuum-Model (RVM) type, while the effective action is also characterised by imaginary parts as well, which imply metastability of this RVM inflation. The axion backgrounds remain undiluted until the exit from inflation, and are responsible for yielding unconventional LV and CPTV Leptogenesis, which is then communicated to the baryon sector (Baryogenesis) via sphaleron processes. The model provides a phenomenologically consistent cosmological evolution from inflation till the modern era.

\end{abstract}

\section{Introduction and Motivation}\label{sec:intro}

\vspace{0.2cm}

The title and contents  of this talk refer to a scenario that links two important concepts in fundamental physics: ``our existence'' and ``gravitational anomaly''. Both need detailed explanations, which will hopefully lead the reader to understand better the claimed link between them. 
Let us commence our analysis by discussing the latter concept first.

\subsection *{Gravitational anomaly} 

\vspace{0.1cm}

If one looks up the term ``{\it gravity anomaly}'' in an encyclopedia or the scientific literature, they will mostly find it to be associated with the difference between the measured value of the gravitational field at a location on Earth and the one predicted by a theoretical model for the same location. The explanations of this phenomenon have to do with the fact that  Earth is not a perfect spheroid, and most importantly with its non-uniform composition, which distorts locally the gravitational field~\cite{GrAn}. Hence such an anomaly  
is an effect of classical gravity.

\begin{figure}[ht!]
    \centering
\includegraphics[width=0.5\textwidth]{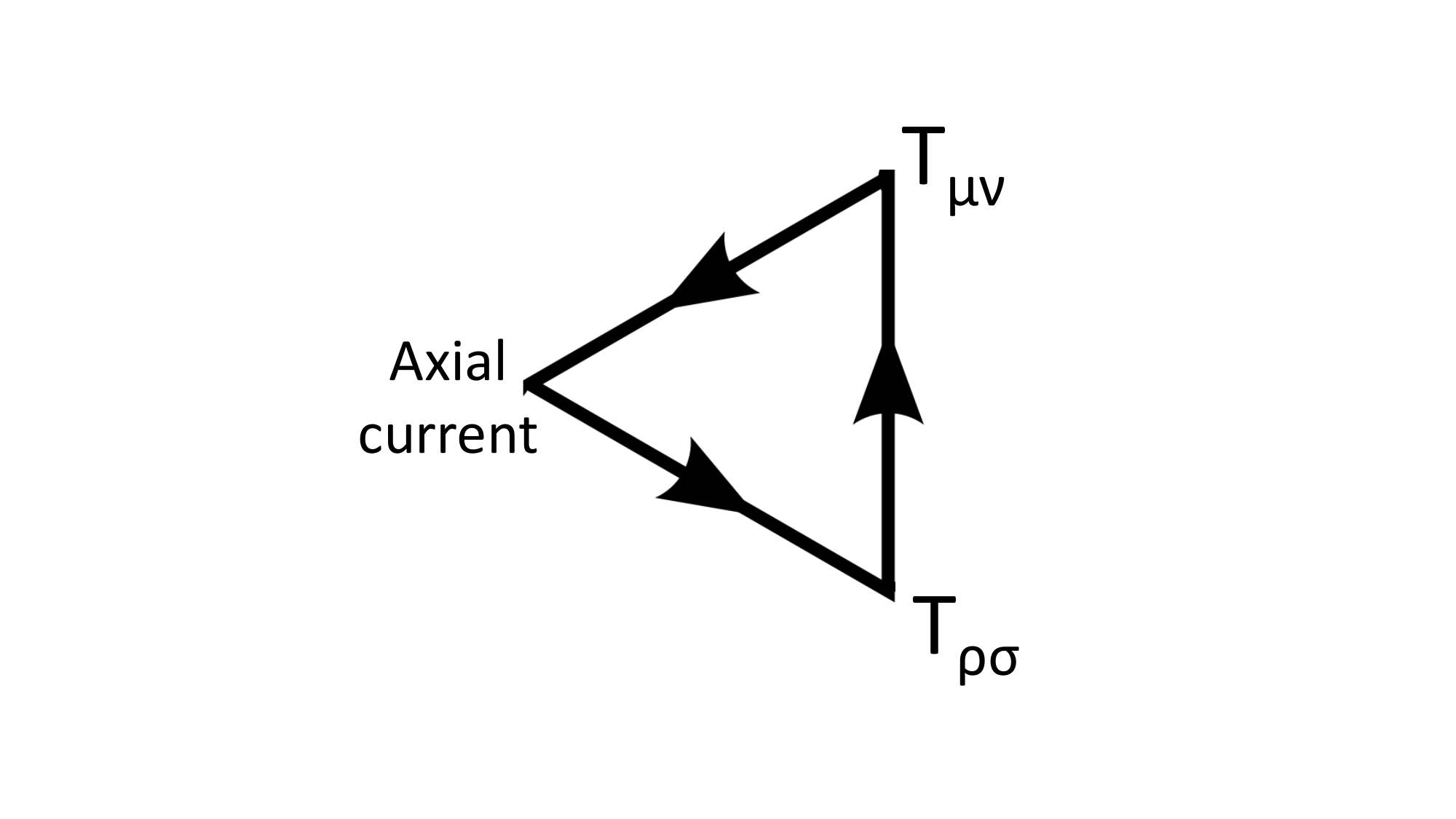}
\vspace{-0.6cm}
    \caption{{\it The triangular gtravitational-anomaly graph in (3+1)-dimensions, with one axial-current insertion, and two insertions of the stress-tensor $T_{\alpha\beta}$. The solid-lines with arrows denote the chiral fermion(s). }}
    \label{fig:triangle}
\end{figure}

However, in theoretical physics, the term ``{\it Gravitational anomaly''} is associated with a quantum phenomenon. In its simplest form, in (3+1)-dimensional spacetimes, it is expressed  by the non-vanishing of a triangular one-loop graph, with three external lines, and three internal lines of chiral-fermions, which  
circulate in the loop (for D-dimensional spacetimes, there are 1+D/2 external lines in the anomalous graph)~\cite{Alvarez-Gaume:1983ihn}. 
It is this concept that we shall be interested in here.
This anomaly, which is a sort of gauge anomaly,
can be best associated with the field theoretical concept of an apparent non-conservation of the stress-energy tensor of a (quantum) matter field theory with chiral fermions, in a curved spacetime.  Naively speaking, this could be interpreted as a breaking of diffeomorphism symmetry for the effective matter theory, that is general covariance. The (3+1)-dimensional-spacetime gravitational anomaly is equivalently linked to the non-conservation of the axial fermion current in the presence of gravitational fields, which is represented by the one-loop diagram of fig.~\ref{fig:triangle},
and consists of chiral-fermions circulating in the loop, and three external lines. One such line accepts an axial-current external insertion, while
the remaining two external lines are associated 
with insertions of the energy-momentum tensor.
In the absence of any gauge fields, the graph of fig.~\ref{fig:triangle} corresponds to the following non-conservation law for the axial fermion current $J_\mu^5 = \sum_{i=1}^{N_f}\overline \psi_i\, \gamma^\mu \, \gamma^5 \, \psi_i $:
\begin{align}\label{dJ5}
J^{5\,\mu}_{\quad \,\,;\mu} = \frac{N_f}{192\, \pi^2} \, R_{CS}\,,
\end{align}
where $;$ denotes gravitational covariant derivative, $N_f$ is the number of flavours of the chiral fermions circulating in the loop, and 
the quantity $R_{CS}$ is the Chern-Simons gravitational anomaly term (or, Hirzebruch signature), defined as:\footnote{Our conventions and definitions used throughout this work are: signature of metric $(-, +,+,+ )$, Riemann Curvature tensor 
$R^\lambda_{\,\,\,\,\mu \nu \sigma} = \partial_\nu \, \Gamma^\lambda_{\,\,\mu\sigma} + \Gamma^\rho_{\,\, \mu\sigma} \, \Gamma^\lambda_{\,\, \rho\nu} - (\nu \leftrightarrow \sigma)$, Ricci tensor $R_{\mu\nu} = R^\lambda_{\,\,\,\,\mu \lambda \nu}$, and Ricci scalar $R_{\mu\nu}g^{\mu\nu}$. We also work in units $\hbar=c=1$.}  
\begin{equation}
	\label{RCS}
	R_{CS}= \frac{1}{2}R^{\mu}_{\,\,\,\nu\rho\sigma}\widetilde{R}^{\nu\,\,\,\,\rho\sigma}_{\,\,\,\mu}\,, \qquad \,\,\,\,\widetilde{R}_{\alpha\beta\gamma\delta}=\frac{1}{2}R_{\alpha\beta}^{\,\,\,\,\,\,\,\,\rho\sigma}\,\varepsilon_{\rho\sigma\gamma\delta}
	\end{equation}
The quantity $\varepsilon_{\mu\nu\rho\sigma}$ is the covariant Levi-Civita totally antisymmetric tensor density, and the  
symbol $\widetilde{(\dots)}$ denotes the gravitationally covariant dual of the Riemann tensor, which should be contrasted with the Hodge-dual~\cite{Eguchi:1980jx} 
\begin{equation}
\label{hodgedualriem}
^\star{R}_{\alpha\beta\gamma\delta}=\frac{1}{2}R_{\alpha\beta}^{\,\,\,\,\,\,\,\,\rho\sigma}\hat \epsilon_{\rho\sigma\gamma\delta}\,,
\end{equation}
which uses the Minkowski-spacetime Levi-Civita symbol $\hat \epsilon_{\mu\nu\rho\sigma}$.
In what follows we shall alternately make use of both duals, exploring the identity 
\begin{align}\label{identity}
\sqrt{-g} \, R_{\mu\nu\rho\sigma} \, \widetilde R^{\mu\nu\rho\sigma} = R_{\mu\nu\rho\sigma} \, ^\star R^{\mu\nu\rho\sigma} \,.
\end{align}
A relevant for our purposes example, in which the quantity $R_{CS}$ is present in the action of the pertinent effective gravitational field theory, is that of
fermionic field theories in contorted geometries~\cite{kaloper}. It suffices for our purposes to consider one flavour ($N_f=1$)  of free Dirac fermions in the presence of torsion~\cite{torsion,torsion2} in the underlying background spacetime. Interactions (e.g. with gauge fields~\cite{kaloper}) will not affect the main conclusions, relevant to our discussion here. 
The corresponding action reads:
\begin{align}\label{axial}
   & \mathcal S_{\rm Torsfermi} = \frac{i}{2} \int d^4x \, \sqrt{-g}\, \Big[ \overline \psi (x) \, \gamma^{\mu} \, \mathcal{D}_\mu (\omega) \, \psi (x) - \overline{\mathcal D_\mu(\omega) \psi(x)} \gamma^{\mu} \, \psi(x) \Big]\, = S_{\rm Dirac-curved}(\mathring{\omega})  \nonumber \\ &+ \frac{1}{8} \, \int d^4x \sqrt{-g} \,  \overline{\psi} (x) \{\gamma^{c}\,, \,
\sigma^{ab}\}\, \mathcal{K}_{abc} \, \psi (x) = S_{\rm Dirac-curved}(\mathring{\omega}) - \frac{3}{4} \int d^4 x\, \sqrt{-g} \, S_\mu \, \overline \psi \, \gamma^{\mu}\, \gamma^{5} \, \psi 
\,,
\end{align}
where $S_{\rm Dirac-curve}(\mathring{\omega})$ is the Dirac Lagrangian of free fermions in the torsion-free curved spacetime with torsion-free spin-connection $\mathring{\omega}$, 
$\mathcal{D}_\mu (\omega)$  is the gravitational covariant derivative, $D_\mu(\omega) \, = \, \partial_\mu + i \, \omega^a_{\,\,\,b\,\mu} \, \sigma{^{b}_{a}}\; , \,\sigma^{ab} \equiv \frac{i}{4} [\gamma^{a}\,, \, \gamma^{b}]$, and the quantity $\omega^a_{\quad \mu \, b} = {\mathring{\omega}}^a_{\quad \mu\, b} + {\mathcal K}^a_{\,\,\,\mu\,b}$ is the contorted spin connection, with $\mathcal K$ the contorsion tensor~\cite{kaloper},
$\mathcal{K}_{abc} =- \frac{1}{2}( T_{cab} - T_{abc} - T_{bca})$, and $T^\mu_{\,\,\,\nu\rho} = - T^\mu_{\,\,\,\rho\nu}$ the torsion.\footnote{Greek indices denote world-indices of the spacetime manifold, $\mathcal M$, while Latin indices pertain to the tangent space of $\mathcal M$ at a spacetime point $x$.} In \eqref{axial} we used the property of the (Dirac) $\gamma$-matrices $\{\gamma^{c}\,, \sigma^{ab}\} = 2 \epsilon^{abc}_{\quad \,d}\, \gamma^{d}\, \gamma^{5}$, 
and we have defined the quantity: 
\begin{align}\label{Sdef}
S_d \equiv \frac{1}{3\!} \epsilon^{abc}_{\quad \, d} \, T_{abc}\,, \qquad \,\,({\rm or~in~form~language} \quad S \equiv \star \,T), 
\end{align}
which is the dual pseudovector constructed out of the totally antisymmetric part of the torsion $T^\mu_{\,\,\,\nu\rho}$. 
From \eqref{axial} we thus observe that only the totally antisymmetric part of the torsion couples to the fermion axial current $J^{5\mu}= \overline \psi \,\gamma^{\mu} \, \gamma^{5} \, \psi$. 

The gravitational (Einstein-Hilbert) part of the action, in the presence of torsion, has the form (where the torsion-$S$ part is written in differential-form language for notational economy reasons):
\begin{align}\label{gravact}
S_{\rm grav} \equiv \frac{1}{2\kappa^2} \int d^4 x \, \sqrt{-g} \Big( R + \widehat \Delta \Big) + \frac{3}{4\kappa^2} \int \, S \, \wedge \, \star \,S \,,
\end{align}
with $\kappa^2 = M_{\rm Pl}^{-2}$ the gravitational constant in 
(3+1)-dimensjons, and $M_{\rm Pl}=2.4 \times 10^{18}$~GeV the reduced Planck mass. The term
$\widehat \Delta$ contains quadratic forms of the remaining  components of the torsion, other than the totally antisymmetric one~\cite{kaloper}.  

Next we consider the total action, which is the sum of \eqref{axial} and \eqref{gravact}: 
\begin{align}\label{stotal}
\mathcal S_{\rm total}  = \mathcal S_{\rm Torsfermi} + S_{\rm grav} \,.
\end{align} 
The classical equations of motion with respect to the (non-propagating) field $S^\mu$ stemming from the action \eqref{stotal}, lead to 
\begin{align}\label{sJ5}
 S = \frac{\kappa^2}{2} J^5 \,.  
\end{align}
If the axial fermion current is conserved, then \eqref{sJ5} would imply~\footnote{The reader should note that the contribution of the symmetric Christoffel symbol to the conservation law of $S^\mu$ vanishes, due to the form of $S^\mu$, \eqref{Sdef}.}
\begin{align}\label{dS}
   {\mathbf d} \star S = 0\,, \qquad ({\rm in~components} \quad \partial_\mu \, S^\mu =0)\,,
\end{align}
where $\mathbf d$ denotes the exterior derivative one form.
Unfortunately, as a result of the gravitational anomaly \eqref{dJ5}, the axial current is not conserved at a quantum level in spacetimes for which $R_{CS} \ne 0$,  for instance, in the presence of chiral gravitational waves (GW)~\cite{Alexander:2004us,Lyth:2005jf}, which will be of interest to us in this talk.

In such cases, to ensure the validity of \eqref{dS}, and thus the conservation of the torsion charge~\cite{kaloper} $Q_s~\equiv~\int \star \,S$, one should add appropriate counterterms to the effective action, order by order in perturbation theory (if we consider, for instance, Quantum Electrodynamics in contorted geometries, as in \cite{kaloper}). In a path-integral approach, this can be implemented by inserting the constraint \eqref{dS}
in the form of a delta functional, 
$\delta (\mathbf d \star S)$,
which can be represented via a pseudoscalar Lagrange multiplier field,\footnote{That the field $\Phi$ is pseudoscalar is disctated by the scalar nature of the $\delta (\mathbf d\, \star\, S)$ term, and the pseudoscalar nature of the argument of the $\delta$ functional.} $\Phi = (\frac{3}{2\kappa^2})^{1/2}\,b$. Thus, the pertinent part of the path-integral (over the $S^\mu$ field) reads
\begin{align}\label{PIS}
    \mathcal Z &\propto \int \mathcal D S \, \exp\Big(i\, \mathcal S_{\rm total}\Big) \, \delta (\mathbf d\star S) \propto
\int \mathcal D S \, \mathcal D b  \,\exp\Big(i\,\int \Big[\frac{3}{4\,\kappa^2}  \, S \, \wedge \, \star \, S - \frac{3}{4} S \, \wedge \, \star j^5 + (\frac{3}{2\kappa^2})^{1/2}\,b\, \mathbf d\,  \star S\Big]\Big) \nonumber \\ &\propto \int \mathcal D b \, \exp\Big(-i \int \Big[\mathbf d b \wedge \star\, \mathbf d b + \frac{1}{192\, \pi^2\, f_b} \, b \, R_{CS}  - \frac{1}{2f_b^2}\, J^5 \wedge \star J^5 \,\Big] \Big) \, , \qquad f_b = \Big(\frac{3\, \kappa^2}{8}\Big)^{-1/2}\,,
\end{align}
where we used \eqref{axial}, \eqref{gravact} (for brevity,  we did not write explicitly the scalar-curvature part of the effective action). In arriving  at the last line, we have 
path-integrated over the non-propagating field $S^\mu$, and integrated by parts the second term in the exponent of the integrand in the second line of \eqref{PIS}, using \eqref{dJ5}.  Thus, we see from \eqref{PIS}, that the effects of the torsion are effectively represented by: (i)  a {\it dynamical} axion-like {\it massless} field $b$ (with canonical kinetic term), which couples to the Chern-Simons gravitational anomaly $R_{CS}$, with an axion coupling $f_b$ given in \eqref{PIS}, and (ii) by the presence of (repulsive) four-fermion terms (quadrature of the axial current). Both these effects are characteristic features of all Einstein-Cartan theories~\cite{Cartan1,Cartan2,Cartan2001RiemannianGI}, although above they have been derived in a relatively simple example, that of free Dirac fermions in contorted geometries. 

A second example, related to the above, where $R_{CS}$ terms coupled to axions appear in the effective action, comes from string theory~\cite{str1,str2,pol1,pol2,kaloper}, and constitutes the main topic of this talk. Indeed, let one consider the Bosonic effective action to ${\mathcal O}(\alpha^\prime)$ (where $\alpha^\prime = M_s^{-2}$, is the Regge slope of the string, with $M_s$ the string mass scale, which in general is different from $M_{\rm Pl}$):
\begin{align}\label{sea4}
    \mathcal S_B =  \int d^4 x \sqrt{-g} \Big(\frac{1}{2\kappa^2}\, R  - \frac{1}{6} H_{\mu\nu\rho}\, H^{\mu\nu\rho} + \dots \Big) \,,
\end{align}
where the $\dots $ denote other terms (gauge, fermionic matter etc.), which we shall not be interested in. Moreover, in \eqref{sea4}, we have assumed that the dilaton has been stabilised to a constant (which we took to be zero, without loss of generality), due to some minimisation procedure, e.g. of appropriate dilaton  potentials induced by string loops~\cite{bms2}. The quantity 
$H_{\mu\nu\rho}$ (totally antisymmetric in its indices) is the field strength of the antisymmetric Kalb-Ramond (KR), spin-one, field $B_{\mu\nu}=-B_{\nu\mu}$ of the massless gravitational multiplet of the closed sector of string theory~\cite{str1,str2}.

Gauge and Gravitational Anomaly cancellation in the extra dimensional spacetime of string theories require the addition of appropriate Green-Schwarz counterterms in the effective action, which amounts to the modification of the definition of the three form $H_{\mu\nu\rho}$ from the simple curl of $B$-field, $\mathbf{d} \, B$, to~\cite{GS}:
\begin{align}\label{csterms}
H &= \mathbf d\,B + \frac{\alpha^\prime}{8\, \kappa} \, \Big(\Omega_{\rm 3L} - \Omega_{\rm 3Y}\Big),  \nonumber \\
\Omega_{\rm 3L} &= \mathring{\omega}^a_{\,\,c} \wedge d\,\mathring{\omega}^c_{\,\,a}
+ \frac{2}{3}  \mathring{\omega}^a_{\,\,c} \wedge  \mathring{\omega}^c_{\,\,d} \wedge \mathring{\omega}^d_{\,\,a},
\quad \Omega_{\rm 3Y} = \mathbf A \wedge  \mathbf d\,\mathbf A + \mathbf A \wedge \mathbf A \wedge \mathbf A,
\end{align}
where, as before, $\mathring{\omega}$ denotes the standard torsion-free spin connection, and $  \mathbf A$ are the non-Abelian gauge fields that generically characterise strings. The $\Omega_{\rm 3L} \, (\Omega_{\rm 3Y})$ are the 
Lorentz (Yang-Mills) CS 3-forms.

The parts of the action \eqref{sea4} which are quadratic in the three form $H$ can be absorbed into a  generalised connection with torsion: 
${\overline \Gamma}_{\mu\nu}^{\rho} = \mathring{\Gamma}_{\mu\nu}^\rho + \frac{\kappa}{\sqrt{3}}\, {\mathcal H}_{\mu\nu}^\rho  \ne {\overline \Gamma}_{\nu\mu}^{\rho}$,
where $\mathring{\Gamma}_{\mu\nu}^\rho = \mathring{\Gamma}_{\nu\mu}^\rho$ is the torsion-free Christoffel symbol.\footnote{This torsion interpretation holds up to and including $\mathcal O({\alpha^\prime}^2)$, {\it i.e.} fourth-order in spacetime derivatives in the effective string-inspired actions~\cite{Metsaev:1987zx,kaloper}.} 
By acting with the exterior derivative on \eqref{csterms} we obtain a Bianchi identity:
\begin{align}\label{bianchi}
   \mathbf d H - \frac{\alpha^\prime}{8\, \kappa} \, {\rm Tr}\Big( R \wedge R - F \wedge F \Big) =0\,.
   \end{align}
The reader should recognise in the second term of the left-hand-side of \eqref{bianchi} the mixed (gravitational and gauge) anomaly term~\cite{Alvarez-Gaume:1983ihn} (the term $R \wedge R$ is proportional to $R_{CS}$ ({\it cf.} \eqref{RCS}),  
in component notation).

The Bianchi identity \eqref{bianchi} is analogous to the conservation of $S$-torsion constraint \eqref{dS}.
Therefore it can be implemented in the $H$-path-integral of the action \eqref{sea4} in a similar way as in the case of Dirac-fermions in contorted spacetimes, examined above, {\it i.e.} via a $\delta$-functional constraint $\delta(\mathbf d  H)$. Path-Integrating $H$, therefore, yields the following effective action, with a (canonically-normalised) dynamical axion field $b$ (called string-model-independent or KR axion~\cite{svrcek}), coupled to the $R_{CS}$~\cite{kaloper,bms} (in component form):
\begin{align}
	S=\int d^4x \,\sqrt{-g}\,  \left[\frac{R}{2\kappa^2}-\frac{1}{2}(\partial_\mu b)(\partial^\mu b) - A\, b\,\Big(R_{CS} + \mathbf F_{\mu\nu} \widetilde{\mathbf F}^{\mu\nu} \Big) + \dots \right]\,, \quad   A=\sqrt{\frac{2}{3}}\frac{\alpha^\prime}{48\kappa} = 
  \sqrt{\frac{2}{3}}\frac{M_{\rm Pl}}{48\, M_s^2}   \,,
\label{eq:Action}  
\end{align}
with $\widetilde{\mathbf F}_{\mu\nu} = \frac{1}{2} \varepsilon_{\mu\nu\alpha\beta}\, \mathbf F^{\alpha\beta}$, the dual of the (non-Abelian) gauge-field strength. 
The model \eqref{eq:Action} constitutes a Chern-Simons gravity~\cite{Jackiw,Alexander:2009tp}, and will be the basis of our discussion in this talk. In fact the axion $b$, and its coupling to the gravitational Chern-Simons (gCS) term $R_{CS}$,
will play a crucial r\^ole 
in ensuring our ``existence'' !
Let us now come to this latter concept. 

\subsection*{Our existence} 
\vspace{0.2cm}

The fact that we exist 
is associated with no other feature of our Universe than simply the fact that 
it is dominated by matter, which hugely overcomes any antimatter component (roughly, today in the Universe, there exists one antiproton in every $10^9$ protons).
Since baryons are the dominant, stable form of ordinary matter in the Universe, this asymmetry between the amounts of matter and antimatter is best expressed as the difference between the densities of baryons ($n_{\rm b})$ and antibaryons ($n_{\overline b})$ in the Universe~\cite{ParticleDataGroup:2024cfk,Planck:2018vyg}:
\begin{align}\label{asymm}
    \eta_{\rm b} \equiv \frac{n_{\rm b} - n_{\overline B}}{n_\gamma} \simeq 6 \times 10^{-10} \times \Big(\frac{\Omega_b \, h^2}{0.022}\Big) \,, \quad {\rm or} \quad {\rm Y}_{\rm B} \equiv \frac{n_{\rm b} - n_{\overline b}}{s}
\end{align}
where $h$ is the reduced Hubble parameter, $\Omega_b$ the baryon abundance in the Universe today, $n_\gamma$ the energy density of photons and $s$ denotes the entropy density of the Universe. On assuming standard cosmology, $s$ is given (in thermal equilibrium) by~\cite{Kolb:1988aj,Kolb:1990vq,Rubakov:2017zvc}: $s \simeq \frac{2\pi^2}{45}\, g_\star \, T^3$, with $T$ the cosmic temperature. The number of degrees of freedom, $g^\star$, depends on the underlying microscopic model. For the Standard Model (SM) of particle physics, $g^\star =\mathcal O(100)$ for $T > 100$~GeV, which is the temperature range corresponding to the decoupling temperature of baryogenesis. The quantity $\Big(\frac{\Omega_b \, h^2}{0.022}\Big) \simeq 1$, according to the Cosmic Microwave Background (CMB) measurements~\cite{Planck:2018vyg}, which yields 
the observed value $\eta^{\rm obs}_{\rm B} \simeq 6 \times 10^{-10}$, in agreement with the Big-Bang-Nucleosynthesis (BBN) data~\cite{Barger:2003zg,ParticleDataGroup:2024cfk}. On the other hand, taking the above expression for $s$ for $T > 100$~GeV, 
yields the observed value of ${\rm Y}_{\rm B} \simeq 0.85 \times 10^{-10}$, for $T > 100$~GeV.

According to the celebrated {\it Sakharov conditions}~\cite{Sakharov:1967dj}, there are four necessary requirements for the emergence of a dominant matter component (over antimatter) in the Cosmos: (i) Violation of charge-parity (CP) Invariance, (ii) Violation of C symmetry, (iii) Violation of Baryon-number, as well as, (iv) deviation from thermal equilibrium, so that the above generated asymmetries are not eventually washed out. Within the SM of particle physics, all these requirements are present~\cite{Kuzmin:1985mm,Rubakov:1996vz,Gavela:1994ds,Gavela:1994dt}, albeit the induced CP Violation is several (about ten) orders of magnitude smaller then the one required to reproduce the observed matter-antimatter asymmetry  \eqref{asymm}.

There is a plethora of important works~(for reviews see, e.g., \cite{Riotto:1999yt,Cline:2006ts}) which attempt to identify extra sources of CP Violation in theories beyond the standard model (BSM), which can lead to the observed baryon asymmetry (Baryogenesis) \eqref{asymm}, such as 
supersymmetry~\cite{Affleck:1984fy,Dine:1995kz,Enqvist:1997si},
electroweak-symmetry-broken-(Higgs)-bubble formation in the electroweak phase transition epoch in Beyond the SM (BSM) extensions (Electroweak Baryogenesis)~\cite{Anderson:1991zb,Cohen:1991iu,Cohen:1993nk,Huet:1994jb,Carena:1996wj,Trodden:1998ym,Morrissey:2012db}, extra-dimensional models and strings (including cosmic strings)~\cite{Casas:1997uk,Cline:1998rc,Allahverdi:2016yws,Davis:1992fm,Alberghi:2010rm,Dolgov:2004ry},
{\it etc.}

\subsection*{The Potential Link of Our Existence to Gravitational Anomalies}
\vspace{0.2cm}

There is, however, another way for the generation of a baryon asymmetry in the Universe. The latter can happen after the creation of a lepton asymmetry (Leptogenesis~\cite{Covi:1996wh,Pilaftsis:2003gt,Giudice:2003jh,Buchmuller:2004nz,Pilaftsis:2005rv,Nardi:2006fx,Abada:2006fw,Davidson:2008bu}) in models which involve massive right-handed neutrinos (RHN)~\cite{Fukugita:1986hr,Luty:1992un,Flanz:1996fb,Pilaftsis:1997jf,Akhmedov:1998qx,Barbieri:1999ma,Buchmuller:2005eh}.
This Baryogenesis-via-Leptogenesis (or Baryoleptogenesis for short)
will be the focus of interest in our review, but from a non conventional point of view entailing Lorentz  and CPT Violating interactions of RHN 
of the type encountered in the
Standard Model Extension (SME)~\cite{kostel,Kostelecky:2008ts}.

At this point we should mention that Sakharov's conditions assume Lorentz and CPT invariance of the effective particle theory responsible for the generation of matter-antimatter asymmetry. If CPT is violated, then the requirement 
of going away from equilibrium in the early Universe may be relaxed. Moreover, with CPT non conservation, one may have direct Baryogenesis (or Leptogenesis) at tree level, in contrast to the conventional CPT invariant theories mentioned above, in which the required CP Violating phases appear at one-loop order~\cite{strumia}.
Direct Baryogenesis in the LV and CPTV SME framework has been discussed in \cite{Bertolami:1996cq}, but in a (phenomenological) context different from the one advocated here.

In our approach, the (spontaneous) Lorentz and CPT Violation will occur as a result of condensation of the gravitational anomaly term $R_{CS}$ in \eqref{eq:Action}, 
that characterises the effective gravitational theories describing the early Universe according to the scenario in \cite{bmsanom,bms,bms2,ms1,ms2,Mavromatos:2022xdo,Dorlis:2024yqw,Dorlis:2024uei}.
These theories belong to the framework of the so-called Running-Vacuum Model (RVM) approach to Modern Cosmology~\cite{rvm1,rvm2,rvmqft1,rvmqft2,rvmqft3,rvmqft4}. The fact that the effective gravitational theories we base our analysis upon in order to  study matter-antimatter asymmetry is derived from strings, leads us to name the model Stringy RVM (StRVM). The Leptogenesis in this approach will characterise effective field theories of RHN in the background of undiluted LV KR-axion backgrounds that are generated as a result of the gCS condensate. The RHN, along with the rest of matter/radiation fields, are created at the era corresponding to the exit from RVM inflation. The resulting effective RHN field theories assume a LV and CPTV SME form~\cite{kostel}.

The structure of the talk is the following: in the next section \ref{sec:model}, we review the cosmological model, and the associated inflation induced by condensation of the gravitational anomalous terms.  
The condensate is created by chiral primordial gravitational waves (GW), in the presence of which, the gravitational anomalous terms (of Chern Simons (CS) type) are non-trivial~\cite{Jackiw,Alexander:2009tp}. The gravitational-CS (gCS) condensate induces a linear KR axion (monodromy) piotential, which drives an RVM-type inflation.
The existence of imaginary parts in the condensate indicate metastability of inflation, whose life time can be estimated in a weak-quantum gravity path integral approach of the GW perturbations about the de Sitter inflationary background. In section \ref{sec:axion}, we discuss the r\^ole of string-model independent KR axions in inducing the necessary CP and CPT Violation for generating a Lepton, and subsequently a Baryon, asymmetry, that is a matter-antimatter asymmetry in this Universe. Indeed,  the gravitational anomaly condensate induces a LV KR axion background, which remains undiluted until the exit from inflation, leading to Leptogenesis during the early radiation epoch, in models with massive right-handed neutrinos. We also discuss briefly the evolution of this KR axion background until the current era, and demonstrate the compatibility of the model with the stringent constraints on LV and CPTV that exist today from a plethora of observations~\cite{Kostelecky:2008ts}. Finally, our conclusions and outlook are given in section \ref{sec:concl}.

\section{The String-inspired Gravitational Model and Condensate Inflation}\label{sec:model}

\vspace{0.2cm}

Our string-inspired gravitational model, which we shall base our analysis upon, is given by the CS gravity model \eqref{eq:Action}, with $\mathbf F=0$, since the basic assumption of refs.~\cite{bmsanom,bms,bms2,ms1,ms2}, where this model was suggested, is that only fields from the gravitational massless multiplet of the string characterise the early Universe field theories as external fields. 

 We take our background spacetime to be that of a spatially-flat, expanding, isotropic, and homogeneous Universe, described by the Friedman-Lemaitre-Robertson-Walker (FLRW) metric:   
\begin{equation}\label{flrwmetric}
    ds^2=-dt^2+\alpha^2(t)\delta_{ij}dx^idx^j\,,
\end{equation}
where $\alpha(t)$ denotes the scale factor. We consider primordial chiral tensor perturbations (GW) of this metric, of the following form: 
\begin{equation}
    ds^2= -dt^2 + \alpha^2(t) (\delta_{ij}+h_{ij})dx^idx^j\; , \quad i,j =1,2,3\,.
\label{FLRWmetricPerturbed}\,,
\end{equation}
where in the so-called helicity (or chiral) basis we have:
\begin{equation}
    h_{ij}(t,\Vec{x})=h_L \ \epsilon_{ij}^{(L)}+h_R \ \epsilon_{ij}^{(R)}=\sum_{\lambda=L,R} h_\lambda(t,\vec{x})\epsilon_{ij}^{(\lambda)}\,,
    \label{helicityexpansion}
\end{equation}
with  $\epsilon_{ij}^{L,R}$  denoting 
the polarisation tensors (for details in the various definitions we refer the reader to the relevant literature 
\cite{Lyth:2005jf,Dorlis:2024yqw}).

Assuming weak GW pertubrations, $|h_{ij}| \ll 1$, we may quantise the graviton fields $h_{ij}$ in the transverse-traceless (TT) gauge: 
$
h^{i}_{\,\,i, j}=0\,, \, h \equiv h_{ij}\,g^{ij}=0$,
where $g^{\mu\nu}$ denotes from now on the background FLRW metric \eqref{flrwmetric}. In \cite{Dorlis:2024yqw,Dorlis:2024uei} we have followed such a canonical quantization of graviton modes, within a weak-quantum-gravity path integral approach in order to evaluate the induced condensate of the gCS term in \eqref{eq:Action}, in the FLRW background \eqref{flrwmetric}. In this approach, we represent averages of composite operators of $\widehat h_{ij}$, $\langle \widehat O_h\rangle$, as path integrals defined with respect to an appropriately defined measure ${\mathcal D} h$, upon TT-gauge fixing the corresponding effective actions that enter the path-integral. For weak tensor perturbations, the effective action stemming from \eqref{eq:Action} is truncated to second order, and thus, the pertinent averages are expressed as~\cite{Dorlis:2024uei}
\begin{equation}\label{Oaver}
  \langle\widehat{\mathcal O}_h\rangle= \frac{1}{\mathcal{Z}}   \int \mathcal{D}h \;\mathcal{\mathcal O}_h\; e^{iS^{(2)}[b,g,h]\Big|_{A=0}-A i \int d^4x \, R_{CS\,h} \, b}\,,    
\end{equation}
where $R_{CS\, h}$ is the perturbed CS anomaly \eqref{RCS} (\eqref{identity}), expanded up to second order in GW perturbations. The partition function $\mathcal Z$ is given by:
\begin{align}\label{ZN}
\mathcal{Z} &=e^{iS^{(0)}[b,g]} \int\prod_{i=1}^{\mathcal N} \mathcal{D}h_i\;\;e^{i\sum_iS^{(2)}[b,g,h_i]\Big|_{A=0}- i \sum_i\int d^4x \;A \, R_{CS\,h_i} \, b} \nonumber \\ &= e^{iS^{(0)}[b,g]} \Big[\int \mathcal{D}h \; e^{iS^{(2)}[b,g,h]\Big|_{A=0}-A i \int d^4x \, R_{CS\,h} \, b}      \Big]^{\mathcal N} \,  .
\end{align}
Above we have assumed $\mathcal N > 1$ dilute (non interacting, to a good approximation) sources of GW~\cite{Dorlis:2024uei}.
A Harteee-Fock (mean field) approximation is applied, according to which one considers {\it chiral} GW perturbations about a constant gCS condensate defined as:\footnote{The chiral ({\it i.e.} left (L), right (R) asymmetric) nature of GW is essential in ensuring the non-triviality of the CS gravitational anomaly terms:
\begin{equation}
R_{\mu\nu\rho\sigma}^{\ \ \ \ \ *}R^{\nu\mu\rho\sigma} =   
   4 i \left(\partial^{2}_z h_L\partial_z h_{R}^{\prime}+  h^{\prime\prime}_L\partial_z h^{\prime}_R -\partial^{2}_z h_R\partial_z h^{\prime}_L-  h^{\prime\prime}_R\partial_z h^{\prime}_L \right)\,,
    \label{RCSconformal}
\end{equation}
where the prime denotes differentiation with respect to the conformal time. The above expression entails the simplification of GW propagation along the $z$ axis. This does not spoil the generality of our arguments in this talk.} 
\begin{align}\label{HF}
\int d^4x \, \mathcal{N}\, R_{CS\, h}\, b =  \int d^4x \sqrt{-g} \, \langle R_{\mu\nu\rho\sigma}\;^*R^{\mu\nu\rho\sigma}\rangle_n \,  b \, + :\int d^4x \, \mathcal{N}\, R_{CS\, h}]
\, b :\,,     
\end{align}
where the symbol $: \dots :$ denotes ({\it normal-ordered}) quantum fluctuations about the condensate, and we used the identity \eqref{identity}, and the definition $\langle R_{\mu\nu\rho\sigma}\;^*R^{\mu\nu\rho\sigma}\rangle_n \equiv n\langle R_{\mu\nu\rho\sigma}\;^*R^{\mu\nu\rho\sigma}\rangle$, with $n= \mathcal N/\sqrt{-g}$.  The reader should  recall the (approximate) constancy of the gCS condensate during inflation, which is confirmed by explicit calculations in the model ~\cite{bms,ms1,ms2,Mavromatos:2022xdo,Dorlis:2024yqw}.
\begin{align}\label{Ncondconst}
\langle R_{\mu\nu\rho\sigma}\;^*R^{\mu\nu\rho\sigma}\rangle_n=~{\rm constant}\,,
\end{align}
during the inflationary era where the Hubble parameter $H \equiv \dot \alpha / \alpha$ 
is approximately constant (with 
the overdot denoting cosmic-time derivative).  The analytic computations of \cite{Dorlis:2024yqw,Dorlis:2024uei}, in which one makes the (plausible) approximation 
that only graviton modes near the UltraViolet (UV) cutoff $\mu = M_s$ of the effective low-energy field theory derived from strings make the dominant contributions,  yield  for the real parts of the gCS during inflation (denoted by the suffix $I$)~\cite{Dorlis:2024yqw}: 
\begin{align}\label{vevRCSfinal}
        {\rm Re}\left<R_{CS}\right>^{\mathcal N_I}_{I} \simeq -\mathcal{N}_I\, \sqrt{\frac{2}{3}} \, \frac{1}{480\, \pi^2}
    \, \Big(\frac{M_s}{M_{\rm Pl}}\Big)^2 \, H_I^4 \, .
\end{align}
The CS condensate leads to inflation self-consistently,\footnote{For details on the pre-inflationary epochs of this StRVM universe, and the passage to the RVM inflation from an axion-dominated stiff era, see \cite{ms1,ms2,Dorlis:2024yqw}.}
as follows from the linear-axion (monodromy) potential that is induced in this case in the action \eqref{eq:Action}, 
$V(b) = A\, b \, {\rm Re}\langle R_{CS}\rangle^{\mathcal N_I}_I$. The fact that the latter leads to a metastable inflationary era, with correct phenomenology, including the estimates of its lifetime,
has been demonstrated by means of a detailed dynamical-system analysis in \cite{Dorlis:2024yqw,Dorlis:2024uei}. The real part of the condensate does not depend on the scheme used for the ordering of the quantum operators $\widehat h_{ij}$ that appear in the CS composite operator $\widehat R_{CS \, h}$ (in contrast to its imaginary parts, see below). The dependence of the gCS condensate on the fourth power of the Hubble parameter leads to this linear-axion(monodromy)-potential-induced inflation being of RVM type~\cite{rvm1,bms,ms1}.

Let us now study the implications of (the real part of) the non-trivial gCS condensate
for the configuration of the KR axion field during inflation.
We commence the discussion by recalling the total derivative nature of the (classical) CS anomaly term: 
\begin{align}\label{totalder}
 R_{CS}= \mathcal J^\mu_{\,\,\,\,\,;\mu}\, ,  
\end{align}
where $;$ denotes covariant derivative in the curved background, and making the following plausible approximation~\cite{bms,ms1} 
in our homogeneous and isotropic FLRW inflationary cosmology with an (approximately) constant  Hubble parameter $H = H_I \simeq {\rm constant}$: 
\begin{align}\label{approx}
\langle R_{CS} \rangle = \langle 
{\mathcal J}^\mu_{\,\,\,\,\,;\mu}\rangle ~ \simeq ~
\partial_0 \langle {\mathcal J}^0 \rangle + 3 \, H_I\, \langle {\mathcal J}^0\rangle\,.  
\end{align}
Then, from the equations of motion for the $b$-field, stemming from \eqref{eq:Action}, we obtain (with $\mathbf F=0$):
\begin{align}
\Box b &= A \langle R_{CS}\rangle \quad \Rightarrow \nonumber \\
{\rm for~FLRW~background}: \quad 
 \ddot b + 3 \, H_I\, \dot b &= A\, \langle R_{CS}\rangle = {\rm constant} \ne 0\,,
\label{eqinflb}
\end{align}
where $\Box$ denotes the covariant D'Alembertian in the FLRW background. 
For $H_I={\rm constant}$, and $\langle R_{CS}\rangle ={\rm constant}$, one obtains from \eqref{eqinflb} that:
\begin{align}\label{bdotconst2}
\dot b = {\rm constant}\,. 
\end{align}
We remark at this juncture that, on writing the equation of motion of $b$, stemming from \eqref{eqinflb}, in the alternative form:
\begin{align}\label{eqinflb2}
\frac{1}{\sqrt{-g}} \, \partial_0 \Big(\sqrt{-g} \, [\dot b -A\, \langle \mathcal J^0 \rangle ] \Big)=0\,,
\end{align}
we may identify the axion background  as $\dot b = \langle \mathcal J^0\rangle$, which is constant, on account of \eqref{bdotconst2}. The dynamical-system analysis of \cite{Dorlis:2024yqw}, for the linear-axion-potential inflation, provides an order of magnitude estimate of $\dot b$:
\begin{equation}
    \dot{b}_I\sim 10^{-1} H_I M_{\rm Pl}\,,
    \label{bdotInflation}
\end{equation}
consistently with the observed phenomenology of inflation~\cite{Planck:2018jri,Planck:2018vyg}. If \eqref{bdotInflation} is integrated over the entire duration of inflation, $\Delta t = t_{\rm end}-t_{\rm onset}$, where $t_{\rm onset}$ ($t_{\rm end}$) denotes the cosmic time at the onset (end) of the RVM inflationary phase, 
then we have 
$b(t_{\rm end}) = b(t_{\rm onset}) + 10^{-1} H_I M_{\rm Pl}\, \Delta t\,, \quad b(t_{\rm onset}) < 0$. Phenomenologically $t_{\rm end}$ is of such an order of magnitude that~\cite{Planck:2018jri,Planck:2018vyg,Martin:2013tda} 
\begin{align}\label{durinfl}
H_I\, \Delta t \sim (50-60)\,,
\end{align}
where the right-hand side of the above relation is the number of e-foldings~\cite{DiMarco:2024yzn}. To ensure consistency with the initial conditions, but also with the de Sitter nature of the gCS condensate term, which should not vary much during the entire duration of inflation, one must have~\cite{bms,ms1,Dorlis:2024yqw}
\begin{align}\label{b0cond}
    |b(t_{\rm onset})| > \mathcal O(10)\, M_{\rm Pl}\,.
\end{align}

On the other hand, the imaginary parts of the condensate 
appear to be quantum operator-ordering dependent~\cite{Dorlis:2024uei}. They are absent in the symmetric (Weyl) ordering, but present in all others. 
The physical operator-ordering scheme to be chosen/fixed should be the one that represents a finite lifetime of inflation, otherwise the quantum theory cannot match the classical dynamical-system approach, where such a finite life time emerges naturally~\cite{Dorlis:2024uei}. 
At this stage, we remind the reader that a quantum-ordering scheme 
among quantum operators $\widehat \phi_i$ is given by  
\begin{align}\label{qord}
{\rm O}_{w_{\rm P}}\Big(\Pi_{i=1}
^N \widehat{\phi}_i\Big) = \sum_{\rm P}\, w_{\rm P}\, \Big(\widehat{\phi}_{i_1} \dots \widehat{\phi}_{i_N}\Big)_{\rm P}\,, \quad \sum_{\rm P} w_{\rm P}=1\,,
\end{align}
where $w_{\rm P}$ denote the weights of the various permutations P of the operators appearing on the right-hand-side of \eqref{qord}. Upon applying 
\eqref{qord} in the evaluation of the composite operator 
corresponding to the gCS term, we obtain:
\begin{align}
\widehat{R_{\mu\nu\rho\sigma}^{\ \ \ \ \ *}R^{\nu\mu\rho\sigma}} = 4 \int \frac{d^{3}\vec{k}d^{3}\vec{k'}}{(2\pi^3)} \ e^{i(\vec{k}+\vec{k'}) \cdot \vec{x}}\, &\Big[ k^2 k' l_{\vec{k'}}\left({\rm O}_{w_{\rm P_1}}\Big(\hat{h}_{L,\vec{k}}\hat{h}^{\prime}_{R,\vec{k'}}\Big) - {\rm O}_{w_{\rm P_2}}\Big(\hat{h}_{R,\vec{k}}\hat{h}^{\prime}_{L,\vec{k'}} 
\Big)\right) \nonumber \\
&- k' l_{\vec{k'}} \left({\rm O}_{w_{\rm P_3}}\Big(\hat{h}^{\prime \prime}_{L,\vec{k}} \hat{h}^{\prime}_{R,\vec{k'}} \Big)
- {\rm O}_{w_{\rm P_4}}\Big(\hat{h}^{\prime \prime}_{R,\vec{k}}\hat{h}^{\prime}_{L,\vec{k'}} \Big) \right) \Big] \ ,
    \label{R*R_quantum}
\end{align} 
where, for the sake of generality, we may assume different ordering schemes (provided by different weights $w_{{\rm P}_i}$, $i=1,\dots, 4$) for each product of operators appearing in 
\eqref{R*R_quantum}, with the weights satisfying 
$\sum_{{{\rm P}_i}=1}^2 \, w_{{\rm P}_i}=1\,, \quad \forall~ \,\, i=1, \dots, 4.$ 

In the simplest scheme, provided by setting only one of the weights, say $w_1$ to one and all others to zero,   we obtain that the imaginary parts of the condensate of the CS anomaly operator ${\rm Im}\langle R_{CS}\rangle$ are given to leading order~\cite{Dorlis:2024uei}: 
\begin{align}
 2 {\rm Im}\langle \widehat{R_{CS}}\rangle \simeq  {\rm Im}\langle \widehat{R_{\mu\nu\rho\sigma}\, \widetilde R^{\nu\mu\rho\sigma}} \rangle = 
 \frac{16  A \text{$\dot{b}$} M_s^7}{7 M^{4}_{\rm Pl} \ \pi ^2} \left[1+ \mathcal O\Big(\frac{H_I^2}{M_s^2}\Big)\right] \ . 
\end{align}
The imaginary parts of the condensate back-react on the effective Lagrangian of GW, and thus on the corresponding Hamiltonian $\mathcal H$, which in this way also acquires an imaginary part~\cite{Mavromatos:2024pho,Dorlis:2024uei}:
\begin{equation}
 {\rm Im}\left(\mathcal H\right) = \int d^3 x \ \sqrt{-g}\,  \frac{1}{2} A \ b \ {\rm Im} \langle \widehat{R_{CS}} \rangle \approx V^{(3)}_{dS} \ \frac{8  b A^2  \text{$\dot{b}$}  M_s^7}{7 M^{4}_{\rm Pl} \ \pi ^2} \, ,
\label{imaginaryhamiltonian}
\end{equation}
where $V^{(3)}_{dS}$ denotes the de Sitter 3-volume of the inflationary spacetime, and the symbol $\approx$ indicates the fact that we keep only leading order contributions in the small quantity $\kappa^2 \dot b$. 
This leads to an estimate of the lifetime $\tau$ of the inflationary vacuum (in natural units, where $\hbar=1$):
$\tau \sim \left(
{\rm Im} \mathcal{H} \right)^{-1}$.
Notably, the metastable nature of inflation of the StRVM is consistent with its embedding in a stringy quantum-gravity framework, as it is consistent with the swampland criteria~\cite{swamp1,swamp2,swamp3,swamp4}.
The (Euclidean) four volume $V^{(3)}_{dS} T^E$ of the (inflationary) de-Siter spacetime, 
with radius $\Lambda =  3 H^{2}_I$, is given by \cite{Fradkin:1983mq}
$
   V^{(3)}_{dS} T^E =  \frac{24 \pi^2}{M^{2}_{\rm Pl} \Lambda}
   $, 
with $T^E$ corresponding to the duration of inflation $\Delta t = \tau$, which phenomenologically is estimated to be of the order indicated in \eqref{durinfl}. 
Then, we obtain:
${\rm Im}\left(\mathcal H\right) = \frac{64 b A^2 \text{$\dot{b}$} \kappa ^6 M_s^7}{ 7 H_I }\cdot\frac{1}{H_I T^E}$.
From \eqref{eq:Action}, \eqref{bdotInflation} and \eqref{b0cond}, on 
leaving the string-scale $M_s$ as a free parameter, we can obtain that the (naively defined) lifetime $\tau = T^E$ of the RVM inflationary vacuum of the StRVM attains the phenomenologically correct value
\eqref{durinfl}, provided:
\begin{equation}\label{MsMPlval}
    H_I\tau = H_I T^E \sim \frac{7H_I^2M_{\rm Pl}^6}{64bA^2\dot{b}M_s^7}\left(H_I T^E\right) \sim 10^{-2}\left(\frac{M_{\rm Pl}}{M_s}\right)^{3} \cdot (H_I T^E)\,\, \Rightarrow \, \, \boxed{\frac{M_s}{M_{\rm Pl}}\lesssim
0.215}\,.
\end{equation}
In the context of the StRVM, this is the decay rate of the unstable RVM ``vacuum'', whose energy is higher than that of the case without the gCS condensate. 
The value of $M_s$ derived in \eqref{MsMPlval} is quite consistent with the findings of the linear-axion-potential dynamical-system analysis of \cite{Dorlis:2024yqw}, which, as already mentioned, fixes this operating-ordering scheme as the physical one in this context.

\section{The r\^ole of KR Axion fields in inducing LV and CPTV Leptogenesis}\label{sec:axion}

\vspace{0.2cm}

The end of inflation in the StRVM framework is characterised by the cancellation of the primordial gCS anomalous terms by the corresponding ones generated by chiral germions that appear as a result of the decay of the running vacuum~\cite{bms,ms1,ms2}. At the end of inflation, therefore, the dynamics of the StRVM is described by the following effective action:
\begin{align}\label{seffferm}
\mathcal S{\rm eff}_{\rm end~infl} = \int d^4 x \, \sqrt{-g} \Big[\frac{1}{2\kappa^2} R  - \frac{1}{2} \partial_\mu b \, \partial^\mu b
- A \, b \, \mathcal J^\mu_{\,\,\,\,;\mu} + \kappa \, b\, J^{5\mu}_{\,\,\,\,\,\,;\mu} \Big] + \dots \,,
\end{align}
where we have taken into account that the primordial gCS terms satisfy the total-derivative relation \eqref{totalder}. The quantity 
$J^{5\mu} = \sum_{i=\rm species} \overline \psi_i \gamma^5 \, \gamma^\mu \, 
\psi $ denotes the chiral current of the various fermion species in the effective theory, generated by the decay of the (false) RVM vacuum, and 
the $\dots$ denote the rest of the terms involving gauge fields and interactions among chiral fermion species. In arriving at \eqref{seffferm} we have also taken into account that the chiral fermions couple to the KR axion in a shift-symmetric way via the chiral anomaly which is given by the covariant gradient of the axial current. In the context of the StRVM, this is also consistent with the interpretation of the (derivative) of the KR axion as being dual to the totally antisymmetric torsion~\cite{kaloper}.

The basic assumption of the scenario of \cite{bms,ms1,ms2}, which we adopted here, is that the number of chiral species circulating in the anomalous (one) loop is such that the last two terms in the right-hand side of \eqref{seffferm} cancel, leaving at most chiral gauge anomalies, either U(1) or non-Abelian QCD type, that is, we impose the relation~\cite{bms2}:\footnote{The cancellation of the gCS anomalies implied by \eqref{anomcanc} means that the leptogenesis scenario of \cite{Alexander:2004us} is not realised in the StRVM. An alternative LV nd CPTV Leptogenesis takes place though~\cite{deCesare:2014dga,Bossingham:2017gtm,Bossingham:2018ivs}, which we shall review in this section.}
\begin{align}\label{anomcanc}
   - A \, b \, \mathcal J^\mu_{\,\,\,\,;\mu} + \kappa \, b\, J^{5\mu}_{\,\,\,\,\,\,;\mu}  = {\rm chiral~U(1)~anomalies~and/or~possibly~QCD-type~chiral~anomalies}\,. 
\end{align}
The QCD anomaly, which is proportional to $G_{\mu\nu}^a \, {\widetilde G}^{a\,\mu\nu}$, $a=1, \dots 8$, where $G_{\mu\nu}^a$ is the gluon tensor, might be generated during the post-inflationary QCD-dominance epoch of the Universe, and lead to instanton-induced potentials, and thus masses, of the KR axion, which thus behaves as a QCD axion~\cite{bms2}. The U(1) chiral anomaly, on the other hand, might be due to cosmic electromagnetic fields in post-inflationary eras, and may persist until the current epoch~\cite{bms}. 

In the scenario of \cite{bms,ms1,ms2}, immediately after exit from inflation, 
the chiral anomalies are not dominant. In that case, 
the equation of motion \eqref{eqinflb2} of the $b$ field from the period at the end of the RVM inflation until Leptogenesis, 
would read (the symbol $\simeq$ indicates equality up to subdominant chiral-anomaly terms):
\begin{align}\label{scale}
 \partial_0 \Big(\sqrt{-g} \, \dot b \Big) \simeq   0\,, 
\end{align}
since in that period $\langle \mathcal J^0 \rangle =0$, due to the aforementioned cancellation of gCS anomalies \eqref{anomcanc}.
Thus, the KR axion field $b$ drops with the temperature of the Universe, as the latter cools down, as $T^3$, 
\begin{align}\label{bscale}
 \dot b(t) = \dot b(t_{\rm end}) \, \Big(\frac{T}{T_H}\Big)^3\,,   
\end{align}
where $T_H = H_I/2\pi$ is the Gibbons-Hawking (observer dependent) temperature of the inflationary de Sitter spacetime~\cite{Gibbons:1977mu}, with $H$ the (approximately) constant value of the Hubble parameter during inflation, 
and 
$\dot b(t_{\rm end})$ denotes the value of the rate of the KR axion background at the end of inflation. In arriving at \eqref{bscale} we took into account that during the radiation era, that succeeds RVM inflation, after the (prolonged, in general~\cite{Lima:2013dmf,Lima:2015mca,Mavromatos:2022yql}) reheating phase, the scale factor of the Universe is inversely proportional to the temperature $T$, $a(t) \sim T^{-1}$.

In view of its approximate constancy during inflation, the value $\dot b(t_{\rm end})$ is of the same order as $\dot b(t_{\rm onset})$, the initial value of the rate of the KR axion field at the onset of the RVM inflation. Hence, from \eqref{bscale}, we finally obtain for the early radiation era, when the U(1) chiral anomalies are assumed absent~\cite{bms}:
\begin{align}\label{bscale2}
    \dot b(t) = 8\pi^3 \, \dot b(t_{\rm onset}) \, \Big(\frac{T}{H_I}\Big)^3\,,
\end{align}
where, as we discussed above, $\dot b(t_{\rm onset})$ satisfies \eqref{bdotInflation} during the entire duration of inflation~\cite{bms,ms1,Dorlis:2024yqw}. On using 
the phenomenologically consistent upper bound for the 
value of the Hubble parameter at inflation~\cite{Planck:2018jri,Planck:2018vyg}
 \begin{align}\label{HI}
 H_I \lesssim  2.5 \times 10^{-5}M_{\rm Pl}\,. 
\end{align}
we can write
\eqref{bscale2} in the form:
\begin{align}\label{bscale3}
  \dot b(t) \sim 4 \times 10^{10} \, M_{\rm Pl}^{-1} \, T^3\,. 
\end{align}

In the scenario of \cite{bms,ms1} this scaling characterises the early radiation era until Leptogenesis 
occurs, in theories with massive RHN, of mass $m$. The relevant Lagrangian density reads:
\begin{align}\label{LRHN}
\mathcal L_{\rm \rm RHN} =  i \overline{ N^c} \, \gamma^\mu \, \partial_\mu \, N  - 
\frac{m}{2} \, \Big(\overline{N^c} \, N + \overline N \, N^c\Big) - \overline{N} \gamma^\mu\, B_\mu \, \gamma^5 \, N - Y_k \, \overline{L_k}\, \widetilde \Phi \, N + {\rm h.c.}\,,
\end{align}
where $N$ denotes the RHN field, assumed to be a Majorana fermion, $L_k$ denotes a lepton doublet in the standard model sector, 
with $k$ a generation index, $\widetilde \Phi$ is the SU(2) dual of the Higgs doublet, and $Y_k \in \mathbb R$, is a (real) Yukawa coupling of the Higgs-portal interaction that connects the RHN sector with the SM sector.
 In our case  $B_\mu$ is related with the $\dot b$ of the KR axion background rate as follows:
\begin{align}\label{B0def}
    B_\mu = M_{\rm Pl}^{-1}\, \dot b \, \delta_{\mu \, 0}\,.
\end{align}
The coupling of $B_\mu$ to the axial RHN current has been dictated by the aforementioned connection of the derivative of the $b$ field, $\partial_\mu b$, to the dual of the torsion~\cite{deCesare:2014dga,bms}. 

As discussed in detail in \cite{Bossingham:2017gtm,Bossingham:2018ivs}, 
for the short period of Leptogenesis, for which the  decoupling temperature\footnote{ The quantity $T_D$ can be estimated by equating the decay rates of the RHN to SM particles with the Hubble rate of the expanding Universe. In the models of \cite{deCesare:2014dga,Bossingham:2017gtm,Bossingham:2018ivs,bms}, which we adopt here, we have $T_D \lesssim m$, for reasons explained in those works, namely the adoption of a delayed decay mechanism~\cite{Weinberg:1979bt}, as in classic Leptogenesis~\cite{Fukugita:1986hr}. In the current work we opt for saturation of the last inequality.} $T_D$ and RHN masses are such that $T_D \lesssim m$, the $T^3$ temperature dependence of $B_0$ \eqref{bscale2} is considered sufficiently mild  so that the (weak, compared to $M_{\rm Pl}^2$) background $B_0 \sim T^3$, is considered as constant, for all practical purposes. In that case,
there is an (analytically computable) asymmetric decay rate between $N \to \phi \, \ell $ and $N \to \overline \phi, \overline \ell$ processes~\cite{deCesare:2014dga,Bossingham:2018ivs,Bossingham:2017gtm}, where $\ell$ denotes SM leptons and the overline the corresponding antileptons, and $\phi$ ($\overline \phi$) the Higgs particle (antiparticle), including charged Higgs particles, which do not decouple as the Leptogenesis in this scenario occurs above the electroweak phase transition, as we shall discuss shortly. Given that $N$ is a Majorana fermion, it is considered its own antiparticle. As noted in \cite{deCesare:2014dga}, a single generation of RHN suffices to produce phenomenologically relevant Leptogenesis, that is a Lepton asymmetry at decoupling $(D)$ of order~\cite{Planck:2018vyg} 
\begin{align}
\frac{\Delta L}{n_\gamma} \Big|_D \, \sim \frac{\Delta L}{s} \Big|_D \, \sim \mathcal O(10^{-10})\,.
\end{align}
 Nonetheless, one may also extend the analysis to incorporate more than one species of RHN, in case one applies the seesaw mechanism~\cite{Minkowski:1977sc,Yanagida:1979gs,Schechter:1980gr,Mohapatra:1979ia,Gell-Mann:1979vob} for generating the phenomenologically relevant mass hierarchies of the light SM neutrinos.

Upon writing
$B_0 (T) = \Phi_0 \, \Big(\frac{T}{m}\Big)^3$, 
it has been shown in \cite{Bossingham:2018ivs}
that the lepton asymmetry generated at the decoupling temperature $T_D \sim m$ in the model \eqref{LRHN} is estimated analytically (using Pad\`e approximant methods) to be (for weak $\dot b \ll M_{\rm Pl}^2$)
\begin{align}\label{leptasym}
\frac{\Delta L (T = T_D \simeq m)}{s} \sim q \, \frac{\Phi_0}{m}\,, \quad q > 0\,,    
\end{align}
where $q$ is a number depending on the details of the freezeout point. In \cite{Bossingham:2018ivs}
$q=\mathcal O(10)$, which we assume below for concreteness. 
From the phenomenological requirement \eqref{leptasym}, and the relations \eqref{bscale3}, \eqref{B0def}, we obtain~\cite{bms}:
\begin{align}\label{mval}
    \frac{m}{M_{\rm Pl}} = \mathcal O\Big(10^{-11}\Big) \sim 10^7 \, {\rm GeV}\,.
\end{align}
From the Higgs-portal interactions in the Lagrangian \eqref{LRHN}, one obtains that, on account of the seesaw mechanism~\cite{Minkowski:1977sc,Yanagida:1979gs,Schechter:1980gr,Mohapatra:1979ia,Gell-Mann:1979vob}, the light (active) neutrinos of the SM sector have masses 
\begin{align}\label{seesaw}
m_\nu \sim |Y_k|^2 v^2 /m\,,    
\end{align}
where $v = \mathcal O(250)$~GeV is the neutral Higgs vev in the elestroweak broken phase of the SM. For phenomenologically relevant upper bound values for the mass of the (lightest) active neutrinos, as implied by oscillation experiments and cosmological data~\cite{Gariazzo:2018pei,Planck:2018vyg}, 
say $m_\nu \lesssim \mathcal O(10)$ meV in the so-called normal hierarchy, we then obtain from \eqref{seesaw} and \eqref{mval}
that~\cite{bms} $|Y_k| \lesssim \mathcal O(10^{-5})$,  which is of similar order to the corresponding Yukawa in the model of unconventional Leptogenesis of ref.~\cite{deCesare:2014dga}. 

Once a lepton number is generated, it can be communicated to the baryon sector via (B-L)-conserving sphaleron processes in the SM sector, as discussed in \cite{Mavromatos:2018map}. In this way, a phenomenologically relevant baryon (and thus matter-antimatter) asymmetry is generated in this StRVM Universe. We hope the reader has appreciated the importance of the KR axion background in the whole process.

After Leptogenesis, the KR-axion background in the StRVM continues to drop with the cosmic temperature as $T^3$, until one reaches a late era during which chiral U(1) anomalies dominate, 
due to cosmic electromagnetic fields in the late Universe~\cite{bms}. At that epoch the scaling of $B_0$ changes to $B_0 \sim T^{2}$~\cite{bms}. Such a scaling, if persistent until today, as in the StRVM~\cite{bms}, would imply a current-era 
value of $B_0^{\rm today} = \mathcal O(10^{-35})$~eV, which lies comfortably within the experimental 
current bounds of the axial vector $B_\mu$ ($B_0 < 10^{-2}$~eV, $B_i < 10^{-22}$~eV, $i=1,2.3$) in models of Lorentz and CPT Violation in the SME framework~\cite{Kostelecky:2008ts}, which the Lagrangian \eqref{LRHN} belongs to, for mildly varying backgrounds $B_0$
with $T$.

\section{Conclusions and Outlook}\label{sec:concl}

\vspace{0.2cm}

In this talk we have presented a cosmological scenario, termed StRVM, which links the important question of the microscopic origin of matter-antimatter asymmetry in the Universe, with the presence of a condensate of gravitational Chern-Simons anomalies, that characterise string-inspired models of inflation within the Stringy Running-Vacuum framework. 
The condensate is induced by chiral primordial gravitational waves, in the presence of which the Chern-Simons anomaly is non trivial. 
The condensate induces a background configuration for the string-model-independent Kalb-Ramond axion $b(x)$ that is present in the model, which is such that the rate $\dot b$ is constant. 
This implies an 
undiluted axion $b(x)$ until the exit from the 
inflationary phase, which in turn results in an unconventional Leptogenesis 
in the early radiation epoch. The latter appears already at the tree-level of the relevant Feynman graphs generating the asymmetry, in contrast to the conventional 
CPT conserving mechanisms, where the Lepton asymmetry is induced at one-loop order. The so-generated Lepton asymmetry can then induce a baryon asymmetry via 
Baryon-minus-Lepton-number conserving sphaleron processes in the standard model sector. 

The presence of the specific axion background implies that the Leptogenesis is of Lorentz- and CPT- Violating type, encountered in effective theories of the Standard Model Extension ~\cite{kostel}. Nonetheless, the model is quite consistent with the current stringent upper bounds of Lorentz and CPT Violation. Indeed, although the LV and CPTV axion background remains non-trivial until the present era, its evolution is such that, as the universe cools down, the axion-background  current-epoch value lies well within the allowed upper bounds for Lorentz and CPT Violation~\cite{Kostelecky:2008ts}.  

We conclude by remarking that, although we presented a StRVM scenario that describes the evolution  of the Universe from inflation until the current era, consistently with the cosmological data, including the generation of matter-antimatter asymmetry, however, it must be noted that the modern era of the StRVM is still not understood fully, especially as far as the microscopic origin of the current-era observed acceleration of the Universe is concerned~\cite{Planck:2018vyg}. As remarked in \cite{bms}, given that ordinary matter has been largely depleted today from the energy budget of the Universe, in favour of Dark energy, it is possible that GW perturbations resurface and dominate the current Universe (provided there is a sufficient number of appropriate sources available in the late Universe). 
This may lead again to the reappearance of a GW-induced condensate of the gCS (which had been cancelled at the end of the RVM inflation, as we have discussed above, by the then generated chiral matter fermions). The condensate is expected (on dimensional grounds) to be proportional to $H_0^2$ terms, where $H_0$ is the present-era Hubble parameter. This is one of the most pressing matters associated with this model that needs to be investigated properly...

\section*{Acknowledgements}
\vspace{0.2cm}

I would like to thank H.T. Elze for the invitation to give a plenary talk in DICE 2024, and for organising such an interesting and thought-stimulating meeting.
This work is supported in part by the UK Science and Technology
Facilities research Council (STFC) and UK Engineering and Physical Sciences Research Council (EPSRC) under
the research grants  ST/X000753/1 and EP/V002821/1, respectively. I also acknowledge participation in the COST Association Actions CA21136 “Addressing observational tensions in cosmology with systematics and fundamental physics (CosmoVerse)” and CA23130 "Bridging high and low energies in search of quantum gravity (BridgeQG)".\\

\bibliographystyle{iopart-num}

\bibliography{bib_GravAnom} 

\end{document}